\documentclass[twocolumn,final,showpacs,amsmath,osajnl]{revtex4}

\usepackage{graphicx}
\usepackage{dcolumn}
\usepackage{bm}
\makeatletter
\def\@dotsep{4.5}
\makeatother

\begin{document}

\title{Diffraction-free beams in thin films}

\author{Carlos J. Zapata-Rodr\'{\i}guez}
\address{Departamento de \'Optica, Universidad de Valencia, Dr. Moliner 50, 46100 Burjassot, Spain.}
\email{carlos.zapata@uv.es}

\author{Juan J. Miret}
\address{Departamento de \'Optica, Universidad de Alicante, P.O. Box 99, Alicante, Spain.}

\begin{abstract}
The propagation and transmission of Bessel beams through nano-layered structures has been discussed recently.
Within this framework we recognize the formation of unguided diffraction-free waves with the spot size approaching and occasionally surpassing the limit of a wavelength when a Bessel beam of any order $n$ is launched onto a thin material slab with grazing incidence.
Based on the plane-wave representation of cylindrical waves, a simple model is introduced providing an exact prescription of the transverse pattern of this type of diffraction-suppressed localized waves. 
Potential applications in surface science are put forward for consideration.
\end{abstract}

%\pacs{050.0050, 130.2790, 260.1180}

\pacs{(OCIS) 260.0260, 310.6870, 350.5500}

\maketitle

\section{\label{sec01} Introduction}

\noindent Nondiffracting Bessel beams are associated with solutions of the Helmholtz wave equation in free space, which transverse amplitude distribution is written in cylindrical coordinates as the product of a linear phase function $\exp \left(i n \theta \right)$ depending on the azimuthal coordinate and a radially-varying Bessel function of the first kind and order $n$ \cite{Stratton41}.
%, where the integer $n$ is the order of the Bessel function\cite{Stratton41}.
The maximum value of the zero-order Bessel function is attained at the origin; however, higher-order Bessel functions have a phase singularity at $r = 0$, reaching its highest intensity nearby \cite{Arfken01}.
Its significance relies on the possibility of engineering laser beams reaching high intensities in a reduced area without diffraction-induced blurring.
In particular, the envelope of the Bessel function falls off as $1/\sqrt{r}$, which leads to an electromagnetic field having an intensity distribution that is not square integrable \cite{Durnin87}.
As a consequence finite-energy practical realizations of Bessel beams show axial bounds between which diffraction-free propagation is valid \cite{Durnin87b,Vasara89,Lu92,Lopez04,Reivelt08}.

The propagation and transmission of Bessel beams through nano-layered structures has been discussed recently \cite{Mugnai01,Longhi04,Zapata08b,Mugnai09} in the context of superluminality in optical wave packets \cite{Mugnai00,Zamboni02,Sheppard02,Porras03,Zapata06d}.
However, an important attribute such as diffraction-free propagation is maintained only if the optical axis of the beam is parallel to the medium interfaces \cite{Manela05,Miret08}.
Guided waves in planar linear \cite{Snyder83} and nonlinear \cite{Horak95} waveguides are of this kind of modal fields having exponentially decaying tails far beyond the core layer.
In general this condition relies on grazing incidence upon the interfaces, which may be used to study surfaces \cite{McGilp95} by increasing wave penetration.

In this paper we conveniently introduce a simple model from which high-order Bessel beams are interpreted as a coherent superposition of propagating and counter-propagating waves along a direction (normal to the layer interfaces).
In fact this sort of cylindrical wave localization is essentially understood as the generation of a line focus around the $z$ axis with $2 \pi$ illumination.
We recognize the formation of diffraction-free waves when a Bessel beam of any order $n$ is launched onto a thin material slab.
The resultant optical beam has in general a twin-peak pattern, which spots size even surpasses the limit of a wavelength.
From a geometrical approach the thin film gives rise to a pair of line images associated with waves impinging onto either the left or the right face of the layer. 

\section{\label{sec02} Vector Bessel beams}

The electric and magnetic fields of a monochromatic diffraction-free wave travelling in an homogeneous medium are written using the modal function ansatz
\begin{subequations}
\begin{eqnarray}
 \mathbf{E} (x,y,z,t) = \mathbf{e} (x,y) \exp \left( i \beta z - i \omega t \right) , \\
 \mathbf{H} (x,y,z,t) = \mathbf{h} (x,y) \exp \left( i \beta z - i \omega t \right) ,
\end{eqnarray}
\label{eq01}
\end{subequations}
where $\omega$ is the time-domain frequency of the radiation and $\beta > 0$ is the propagation constant along the $z$ axis.
We consider a non-absorbing medium with positive dielectric constant ($\epsilon > 0$), zero magnetic susceptibility, and free of charges.
Let us assume a polarized wave such that the $y$-component of the electric field vanishes, $E_y = 0$; complementary polarized waves with $H_y = 0$ may be obtained straightforwardly by substituting $\mu_0 \leftrightarrow \epsilon_0 \epsilon$, $\mathbf{E} \to \mathbf{H}$, and $\mathbf{H} \to - \mathbf{E}$.
Faraday's law leads to the estimation of the magnetic field in terms of the electric field,
$\mathbf{H} = \left( -i / \omega \mu_0 \right) \nabla \times \mathbf{E}$.
Furthermore we obtain the equation $E_z = i \beta^{-1} \left( \partial E_x / \partial x \right)$ from the Coulomb's law.
Therefore we may use the two-dimensional scalar function $U(x,y) \equiv e_x$ in order to describe the wave mode unambiguously.
Let us point out that alternate routes for the description of electromagnetic diffraction-free beams may be found elsewhere \cite{Bouchal95b,Bouchal98,Paakkonen02}.

\begin{figure}
\centering
\includegraphics[width=8cm]{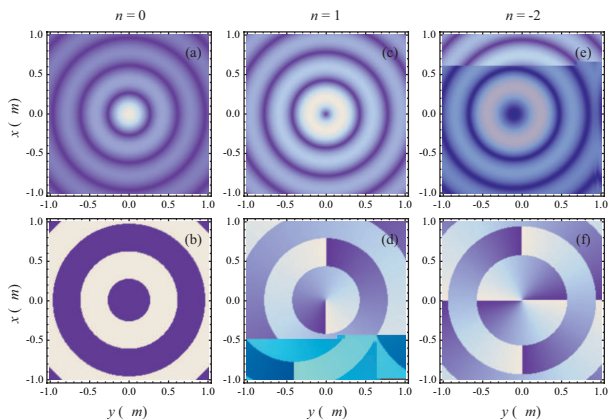}
\caption{Normalized amplitude (upper row) and complex argument (lower row) of the wave function $U_n (x,y)$ for $\kappa = 9.07~\mu \mathrm{m}^{-1}$ and indices (a-b) $n = 0$, (c-d) $n = 1$, and (e-f) $n = -2$.} 
\label{fig01}
\end{figure}

The function $U(x,y)$ satisfies the differential wave equation
\begin{equation}
 \frac{\partial^2 U}{\partial x^2} + \frac{\partial^2 U}{\partial y^2} + \left( k^2 - \beta^2 \right) U = 0 ,
\label{eq03}
\end{equation}
where the wavenumber $k = \omega \sqrt{\epsilon} / c$ and $c$ is the speed of light in vacuum.
Particular solutions of the wave equation (\ref{eq03}) that are periodic in the azimuthal coordinate $\theta$, $U_n = f_n (r) \exp \left( i n \theta \right)$ being $r$ the radial coordinate and $n$ an integer, leads the radial function $f_n$ to satisfy the Bessel's equation \cite{Stratton41}.
Assuming $f_n(0) < \infty$ and $\beta < k$ we may write $f_n (r) = J_n (\kappa r)$, where $J_n$ is a Bessel function of the first kind and $\kappa^2 = k^2 - \beta^2$.
In Fig.~\ref{fig01} we plot the 2D wave field $U_n$ of different orders $n$ at a wavelength $\lambda = 0.6\ {\mu m}$ ($k = 2 \pi / \lambda$) and propagation constant $\beta = k/2$ ($\kappa = \sqrt{3} \beta$).

\section{\label{sec03} Radiation modes with on-axis localization inside a thin layer}

In this paper we consider the monochromatic nondiffracting beam is propagated in the layered medium of Fig.~\ref{fig02}(a).
The $y$ axis is set such that it is perpendicular to the surfaces separating the centered slab (of width $L$) with dielectric constant 
\begin{equation}
 \epsilon_c = \epsilon \left( 1 + \delta \right),
\end{equation}
and the cladding semi-infinite layers of dielectric constant $\epsilon$.
Therefore, $E_y = 0$ ($H_y = 0$) indicates that the diffraction-free beam is s-polarized (p-polarized).
The existence of the Bessel beam is assumed in the region $|y| > L/2$.
In the limit $L \to 0$ it obviously yields a radially-symmetric perfect Bessel beam.
Our purpose is the analysis of the Bessel-beam transformation in the presence of the thin film.
For the sake of clarity in our formulation, let us first consider the limiting case setting either $L = 0$ or $\delta = 0$ (perfect match, $\epsilon_c = \epsilon$).

\subsection{\label{sec03a} Propagating and counter-propagating modal decomposition}

\begin{figure}
\centering
\includegraphics[width=7cm]{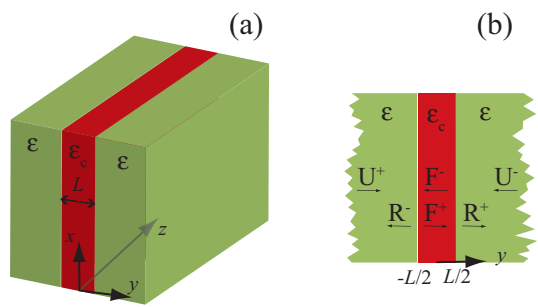}
\caption{(a) Schematic geometry of the planar-layer-based medium.
 In (b) we represent the wave fields in regions established by interfaces at $y = \pm L/2$.
 Perfect matching $\epsilon_c = \epsilon$ leads to wave fields $U^\pm = F^\pm = R^\pm$.} 
\label{fig02}
\end{figure}

From a practical point of view it is necessary to describe the wave field by means of propagating and counter-propagating fields, $U_n = U_n^+ + U_n^-$.
The propagating field $U_n^+$ is assumed to be excited at $y \to -\infty$ traversing the semi-infinite layer in the positive direction of the $y$ axis, as depicted in Fig.~\ref{fig02}(b).
% and impinging on the left facet of the slab, $y = - L/2$. 
On the other hand, $U_n^-$ emerges from $y \to + \infty$ to go through the right-side cladding layer directed in the negative direction. 
One can see that an approach describing $U_n$ in rectangular coordinates is preferable.

To obtain a convenient representation of the fields $U_n^+$ and $U_n^-$, let us recall the Sommerfeld's integral representation of the Bessel functions \cite{Stratton41},
\begin{equation}
 J_n \left( \rho \right) = \frac{i^{-n}}{2 \pi} \int_{-\pi}^{\pi} \exp \left( i \rho \cos \phi + i n \phi \right) d \phi .
\label{eq04}
\end{equation}
Consequently, a Bessel beam can be cast into homogeneous plane waves whose wave vector directions form a circular cone around the $z$ axis \cite{Indebetouw89},
\begin{equation}
 U_n = \frac{i^{-n}}{2 \pi} \left( \int_{-\pi}^0 + \int_0^\pi \right) \exp \left[ i n \alpha + i \kappa \left( x \cos \alpha + y \sin \alpha \right) \right] d \alpha .
\label{eq05}
\end{equation}
Denoting $\gamma$ the angle measuring the aperture of the cone we obtain $\kappa = k \sin \gamma$ and $\beta = k \cos \gamma$.
In Fig.~\ref{fig01} $\gamma = 60\ \mathrm{deg}$.
When $0 < \alpha < \pi$, the projection of the wave vector onto the $y$-axis $\kappa \sin \alpha$ is positive, representing a wave field whose phase advances toward positive values of $y$; this shall be written as $U_n^+$.
Otherwise $\kappa \sin \alpha < 0$ for $-\pi < \alpha < 0$ leading to a counter-propagating field, $U_n^-$. 
Setting $k_x = \kappa \cos \alpha$ in Eq.~(\ref{eq05}) we obtain a 1D Fourier representation of the field components
%\begin{equation}
% U_n^\pm (x,y) = \frac{1}{2 \pi} \int_{-\kappa}^{\kappa} \tilde{U}_n^\pm (k_x,y) \exp \left( i k_x x \right) d k_x ,
%\label{eq06}
%\end{equation}
\begin{equation}
 U_n^\pm (x,y) = \frac{1}{2 \pi} \int_{-\kappa}^{\kappa} \tilde{U}_n^\pm (k_x)  \exp \left( \pm i k_y y \right) \exp \left( i k_x x \right) d k_x ,
\label{eq06}
\end{equation}
where the spatial spectrum 
%$\tilde{U}_n^\pm (k_x,y) = \tilde{U}_n^\pm (k_x)  \exp \left( \pm i k_y y \right)$,
\begin{equation}
 \tilde{U}_n^\pm (k_x) = i^{-n} \frac{\left( k_x \pm i k_y \right)^n}{k_y \kappa^n} ,
\label{eq07}
\end{equation}
and $k_y = \sqrt{\kappa^2 - k_x^2}$.
%Therefore the spectrum of the propagating field $a^+$ employs $+$ and the counterpropagating term $a^-$ uses $-$.

We point out that the cylindrical wave $U_0$ given in Eq.~(\ref{eq05}) for $n = 0$ may be essentially understood as an aberration-free 2D focused field with focus at the origin, $x = y = 0$, using $2 \pi$ illumination.
Setting $n \alpha = k \Phi$ allows us to identify the aberration function \cite{Born99} $\Phi$ that would depend linearly upon the Bessel index $n$ and the azimuthal coordinate $\alpha$.

\begin{figure}
\centering
\includegraphics[width=8cm]{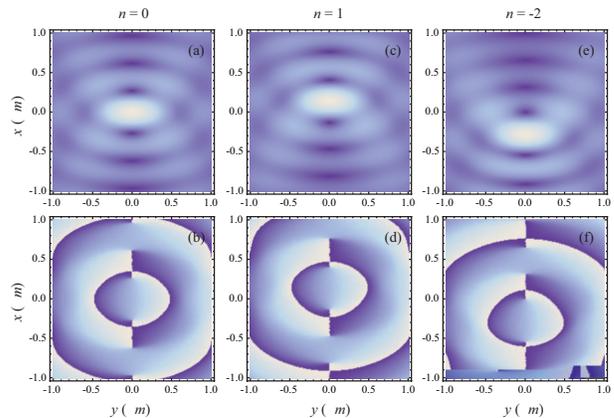}
\caption{Wave function $U_n^+ (x,y)$ associated with Bessel beams of Fig.~\ref{fig01}.
 Similarly, normalized amplitude is represented in the upper row and complex argument is plotted in the lower row.} 
\label{fig03}
\end{figure}

Let us conclude our analysis on $\tilde{U}_n^\pm$ by giving some functional expressions for its fast computation.
For nonabsorbing dielectric media ($\epsilon > 0$) we obtain the relation $\tilde{U}_n^- = (-1)^n (\tilde{U}_n^+)^*$.
This means that
\begin{equation}
 U_n^- (x,y) = (-1)^n \left[ U_n^+ (-x,y) \right]^* ,
\end{equation}
so that propagating and counter-propagating components of the Bessel wave functions possess mirror symmetry with respect to the plane $x = 0$.
Furthermore, using the Jacobi-Anger expansion \cite{Arfken01} in Eq.~(\ref{eq05})
%\begin{equation}
% \exp \left( i z \cos \theta \right) = \sum_{m = - \infty}^\infty i^m J_m (z) \exp \left( i m \theta \right) ,
%\end{equation}
we may obtain the analytical expression
\begin{equation}
 U_n^+ (x,y) = \frac{1}{2} \sum_{m = - \infty}^\infty \mathrm{sinc} \left[ (m-n)/2 \right] U_m (x,y) ,
\end{equation}
where the function $\mathrm{sinc} (u) = \sin(\pi u) / (\pi u)$ and $\mathrm{sinc} (0) = 1$.
%\begin{equation}
% U_n^+ = \frac{U_n}{2} + \sum_{m \neq - n} b_m^{(n)} J_m (\kappa r) \exp \left( - i m \theta \right) ,
%\end{equation}
%where
%\begin{equation}
% b_m^{(n)} = \frac{(-1)^m}{\pi (n + m)} \sin \left[ (n+m) \pi / 2 \right] .
%\end{equation}
In Fig.~\ref{fig03} we represent $U_n^+ (x,y)$ associated with the Bessel beams shown in Fig.~\ref{fig01}.
Radial symmetry characteristic of Bessel patterns is not conserved here, exhibiting an elongated peak along the $y$ axis.
Moreover, this \emph{focus} is shifted toward positive (negative) values of the $x$ coordinate for $n > 0$ ($n < 0$).
Finally, the on-axis vortex \cite{Soskin97} for the order $n = -2$ is maintained (not for $n = 1$); however in Fig.~\ref{fig03}(f) we observe that the topological charge has changed.
Additional phase singularities come out at points all along the $x$ axis even for $n = 0$.

\subsection{\label{sec03b} Transverse wave fields}

Let us now evaluate the field transformation of the Bessel beam $U_n$ in the presence of the core layer.
We assume a wave field $U_n^+$ [see Eqs.~(\ref{eq06}) and (\ref{eq07})] propagating in the semi-infinite medium $y < -L/2$ and impinging at the core-cladding left interface, as seen in Fig.~\ref{fig02}(b).
Also a counter-propagating field $U_n^-$ evolving in the semi-infinite medium $y > L/2$ is incident at the right face of the core layer.
In general we may represent the $x$-component of the electric field transmitted in $|y| < L/2$ employing a factorization of the form given in Eqs.~(\ref{eq01}).
Also separating the propagating and counter-propagating parts of the field as $F_n^+ + F_n^-$ we may use the Fourier integral representation shown in Eq.~(\ref{eq06}) substituting $k_y$ by $k_{c,y} = \sqrt{k_c^2 - \beta^2 - k_x^2}$, where $k_c = \omega \sqrt{\epsilon_c} / c$.
Explicitly we write
\begin{equation}
 F_n^\pm (x,y) = \frac{1}{2 \pi} \int_{-\kappa}^{\kappa} \tilde{F}_n^\pm (k_x)  \exp \left( \pm i k_{c,y} y \right) \exp \left( i k_x x \right) d k_x .
\label{eq12}
\end{equation}
Here $k_{c,y} = k'_{c,y} + i k''_{c,y}$ is in general a complex number, where $k'_{c,y} \ge 0$; also $k''_{c,y} \ge 0$ leading to either homogeneous or damped waves inside the slab.

Additionally, the fields $R_n^-$ (in $y < -L/2$) and $R_n^+$ (in $y > L/2$) flying away from the core layer should be included for a complete description, schematically shown in Fig.~\ref{fig02}(b).
In core barriers, these fields account for reflected waves from $U_n^+$ and $U_n^-$, respectively.
However, in the limit $\delta \to 0$ we would have $R_n^\pm = U_n^\pm$ related to transmitted waves through the core medium.
In general, 
\begin{equation}
 R_n^\pm (x,y) = \frac{1}{2 \pi} \int_{-\kappa}^{\kappa} \tilde{R}_n^\pm (k_x)  \exp \left( \pm i k_y y \right) \exp \left( i k_x x \right) d k_x ,
\end{equation}
stand for a combination of reflected and transmitted waves as shown below.

We point out that the condition $\beta < k$ get rid of a diverging behavior of $U_n^\pm$ far off the core layer.
These onedirectional wave fields on other hand lead to the formation of localized radiation modes.
Such an optical feed is necessary to obtain a real-valued propagation constant.
Otherwise it gives rise to leaky modes \cite{Hu09} being $\beta$ a complex number, which might be understood as disability to suppress the natural intensity decay induced by diffraction.

In the following, the spatial spectra $\tilde{F}_n^\pm$ and $\tilde{R}_n^\pm$ are estimated.
%It is obvious that perfect matching $\epsilon_c = \epsilon$ leads to wave fields $U_n^\pm = F_n^\pm = R_n^\pm$, which is analyzed above.
Boundary conditions for the s-polarized waves at the core-cladding interfaces lead to the continuity of the scalar field $E_x$ and its normal derivative $\mu_0^{-1} \partial E_x / \partial y$ at $y = \pm L/2$.
Therefore the propagation constant $\beta$ is conserved in the layered media.
Also we obtain
%\begin{widetext}
%\begin{subequations}
%\begin{eqnarray}
% \tilde{R}_n^\pm (k_x) = - \frac{2 i \exp[ i (k_{c,y} - k_y) L] \left[ -2 i k_{c,y} k_y \tilde{U}_n^\pm + \left( k_{c,y}^2 - k_y^2 \right) \tilde{U}_n^\mp \sin (k_{c,y} L) \right]}{\exp (2 i k_{c,y} L) \left( k_{c,y} - k_y \right)^2 - \left( k_{c,y} + k_y \right)^2} , \\
% \tilde{F}_n^\pm (k_x) = - \frac{2 \exp[ i (k_{c,y} - k_y) L / 2] k_y \left[ \exp (i k_{c,y} L) (k_{c,y} - k_y) \tilde{U}_n^\mp + \left( k_{c,y} + k_y \right) \tilde{U}_n^\pm \right]}{\exp (2 i k_{c,y} L) \left( k_{c,y} - k_y \right)^2 - \left( k_{c,y} + k_y \right)^2} .
%\end{eqnarray}
%\label{eq08}
%\end{subequations}
%\end{widetext}
\begin{equation}
 \left[
 \begin{array}{c}
  \tilde{R}_n^\pm \\
  \tilde{F}_n^\pm
 \end{array}
 \right]
 = 
 \left[
 \begin{array}{cc}
  A & B \\
  C & D
 \end{array}
 \right]
 \left[
 \begin{array}{c}
  \tilde{U}_n^\pm \\
  \tilde{U}_n^\mp
 \end{array}
 \right] ,
 \label{eq10}
\end{equation}
where
\begin{subequations}
\begin{eqnarray}
 A &=& \frac{4 k_{c,y} k_y \exp[ i (k_{c,y} - k_y) L]}{\left( k_{c,y} + k_y \right)^2 - \left( k_{c,y} - k_y \right)^2 \exp (2 i k_{c,y} L)} , \\
 B &=& \frac{2 i \left( k_{c,y}^2 - k_y^2 \right) \sin (k_{c,y} L) \exp[ i (k_{c,y} - k_y) L]}{\left( k_{c,y} + k_y \right)^2 - \left( k_{c,y} - k_y \right)^2 \exp (2 i k_{c,y} L)} , \\
 C &=& \frac{2 k_y \left( k_{c,y} + k_y \right) \exp[ i (k_{c,y} - k_y) L / 2]}{\left( k_{c,y} + k_y \right)^2 - \left( k_{c,y} - k_y \right)^2 \exp (2 i k_{c,y} L)} , \\
 D &=& \frac{2 k_y (k_{c,y} - k_y) \exp (i k_{c,y} L) \exp[ i (k_{c,y} - k_y) L / 2]}{\left( k_{c,y} + k_y \right)^2 - \left( k_{c,y} - k_y \right)^2 \exp (2 i k_{c,y} L)} .
\end{eqnarray}
\label{eq09}
\end{subequations}
Excepting the phase factor $\exp \left( -i k_y L \right)$, the terms $A(k_x)$ and $B(k_x)$ are simply the Airy's formulas corresponding to the transmittance and reflectance coefficients through the slab \cite{Yeh88}.
Thus, the matrix elements $A$ and $B$ quantify the contribution of reflected and transmitted wavelets to the scattered fields $R_n^\pm$.

We confirm that $\tilde{U}_n^\pm = \tilde{F}_n^\pm = \tilde{R}_n^\pm$ under perfect matching conditions; note that $\delta = 0$ yields $k_{c,y} = k_y$.
However, serious deviations arise when $|k_{c,y} - k_y| L \ge 1$ for which the phase (exponential) factor in the numerator of the coefficients $ABCD$ given in Eqs.~(\ref{eq09}) oscillates (falls off) rapidly.
This inequality is approximately satisfied if $|\delta| L \ge 2 \kappa / k^2$.
Specifically for both material and beam sizes in the order of the wavelength, $L \sim \lambda$ and $\kappa \sim k$, a case that is analyzed in the next section, it yields $|\delta| \ge 1 / \pi$ so that the difference of refractive indices for the core and the cladding layers might not be large in excess.

\section{\label{sec04} Numerical simulations}

Let us first evaluate numerically the amplitude distribution of the wave function $e_x$ corresponding to Bessel excitations as those of Fig.~\ref{fig01} in a centered dielectric slab of dielectric constant with relative increment $\delta = 10^{-2}$.
Therefore the source spectra $\tilde{U}_n^\pm$ are those of Eq.~(\ref{eq07}).
We choose $L = 100\ {\mu m}$ in order to hold the inequality $|\delta| L \ge (\sin \gamma) \lambda / \pi$ at $\lambda = 0.6\ {\mu m}$ and $\gamma = 60\ \mathrm{deg}$.
In this case the slab width is much larger than the spot size of the Bessel source so that wave localization is expected at points far from the surfaces. 

\begin{figure}
\centering
\includegraphics[width=8cm]{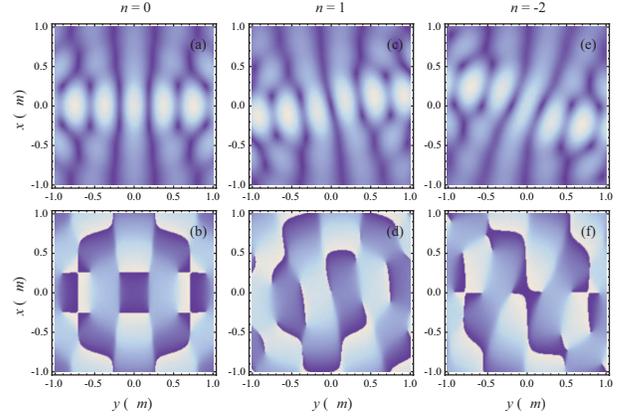}
\caption{Transverse distribution of Bessel-excited wave field $e_x (x,y)$ inside a thin slab of width $L = 100\ {\mu m}$ made of a material with relative dielectric constant $\delta = 10^{-2}$.
 Again $|e_x|$ is plotted in the upper row and its complex argument is shown in the lower row.} 
\label{fig04}
\end{figure}

The resultant wave function,
\begin{equation}
 e_x (x,y) = 
  \left\{ 
   \begin{array}{ll}
    U_n^+ (x,y) + R_n^- (x,y) & \mbox{if $y < - L/2$};\\
    F_n^+ (x,y) + F_n^- (x,y) & \mbox{if $|y| < L/2$};\\
    U_n^- (x,y) + R_n^+ (x,y) & \mbox{if $y > L/2$},
   \end{array} 
  \right.
 \label{eq11}
\end{equation}
is depicted in Fig.~\ref{fig04}.
Irrespective of the Bessel index $n$, the central high-field patterns shown in subfigures from the upper row consist of twin spots with large sidelobes.
For $n = 0$ the spots appear along the $y$ axis but an additional lateral displacement may be observed in the transverse $x$ direction if $|n|>0$.
Also a plethora of phase dislocations are found in all cases.

\begin{figure}
\centering
\includegraphics[width=6cm]{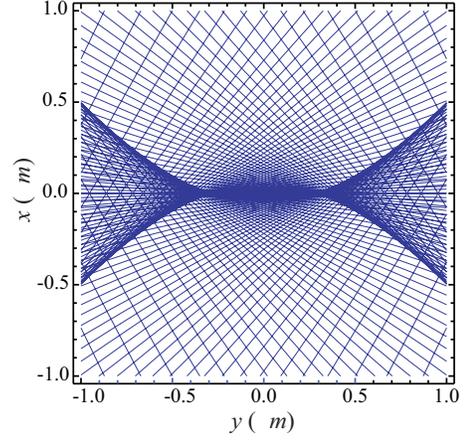}
\caption{Ray tracing representation of the Bessel-like beams shown in Fig.~\ref{fig04}.
 Dual parallel axes at $y = \pm 0.25\ {\mu m}$ (and $x = 0$) set the confinement region of the nondiffracting beam.
 Spherical aberration drawing the envelope of the refracted rays gives rise to a bilateral-symmetric caustic.} 
\label{fig04prev}
\end{figure}

The geometrical approach ($\lambda \to 0$) bring us a simple interpretation for the results of this simulation:
We may think of the incident focal wave $U_n^+$ as a 2D system of homocentric rays, which common point (focus) is at the origin $(x,y) = (0,0)$. 
These rays are refracted at the plane surface $y = - L/2$ becoming into the field $F_n^+$, for which the focus is paraxially imaged at $y = (\sqrt{\epsilon_c / \epsilon} - 1 ) L/2 \approx \mathrm{Re} (\delta) L / 4$.
In our numerical example, the paraxial image would be placed at $y = 0.25\ {\mu m}$.
Also it would exhibit strong spherical aberration.
Applying a similar picture to the counter-propagating wave field $U_n^-$ we are able to recognize the optical ray tracing shown in Fig.~\ref{fig04prev} for the region near the $z$ axis.
In the plot we represent for simplicity the projection of ray traces onto the $xy$ plane inside the lamellar dielectric structure.
Our previous discussion supports the high ray concentration that is produced around the dual optical axes $(x,y) = (0,\pm 0.25)\ {\mu m}$.
We may conclude that Bessel excitations in the presence of the thin dielectric layer yield a pair of localized wave fields whose proximity leads to a linear interaction still conserving the diffraction-free characteristic of the originating beam.

%Conversely, strong wave impendance mismatching ($|\delta| \to \infty$) leads to the limiting values $\tilde{F}_n^\pm = 0$ and $\tilde{R}_n^\pm = - \exp( - i k_y L) \tilde{U}_n^\mp$.
In order to analyze surface effects we reduce the slab width $L$ approaching a wavelength and, additionally, the wave impedance mismatching is increased.
%A convenient procedure consists in leaving the product $|\delta| L$ maintained unaltered. 
In the limit $|\delta| \to \infty$ we obtain $\tilde{F}_n^\pm = 0$ and $\tilde{R}_n^\pm = - \exp( - i k_y L) \tilde{U}_n^\mp$ from Eqs.~(\ref{eq10}) and (\ref{eq09}).
In this case, the core layer behaves like a infinite tunneling barrier prohibiting the transit of the wave through itself.
Thus each reflected field,
\begin{equation}
 R_n^\pm (x,y) = - U_n^\mp (x,-y \pm L) ,
\end{equation}
reproduces a mirror-symmetric version of its corresponding incident wave field in the cladding medium.
Also the phase factor of argument $- k_y L$ in the spatial spectrum leads to a displacement $L$ along the $y$ axis of the original focus.
That is, the geometric focus of the wave field $U^\pm$ is stigmatically imaged by means of $R^\mp$ at $(x,y) = (0,\mp L)$, respectively.

\begin{figure}
\centering
\includegraphics[width=8cm]{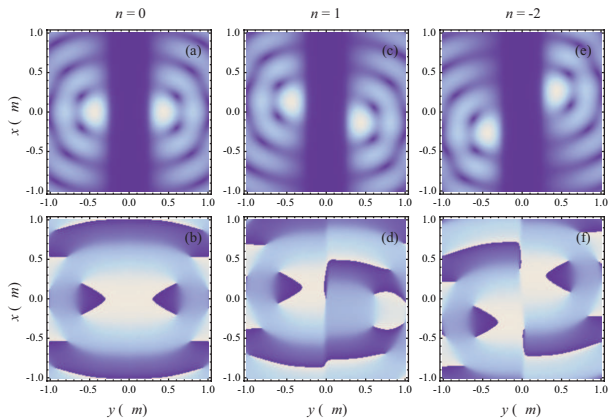}
\caption{Resultant field $e_x (x,y)$ for different Bessel indices $n$ in the vicinities of a thin slab with $L = 0.5\ {\mu m}$ and $\delta = - 10$.
 Subfigures in the upper row account for amplitudes; associated complex arguments are depicted below.} 
\label{fig05}
\end{figure}

Fig.~\ref{fig05} shows the complex amplitude and phase of the wave function $e_x (x,y)$ given in Eq.~(\ref{eq11}) for material parameters $L = 0.5\ {\mu m}$ and $\delta = - 10$.
A bifocal 2D pattern is clear in the amplitude figures following a previous discussion.
Interestingly, the size of the twin spots is significantly lower in the $y$ direction than that of the source fields $U_n^\pm$, and it is close to the width of the $J_0$ pattern.
However, this superresolving effect is nearly lost if $L$ increases sufficiently.
We may conclude that the interference of propagating and counter-propagating fields, whose spectra approaches $\tilde{U}_n^+$ in $y < -L/2$ and $\tilde{U}_n^-$ in $y > L/2$ in the limit $|\delta| \to \infty$, yields a couple of intense superresolving spots near the surfaces of the thin core film.
This effect is independent of the Bessel index $n$, however it drives a lateral shift along the $x$ axis.
Finally, one can see in Fig.~\ref{fig05} a set of phase singularities, which charge is independent of the index $n$ from the Besselian source.

\begin{figure}
\centering
\includegraphics[width=8cm]{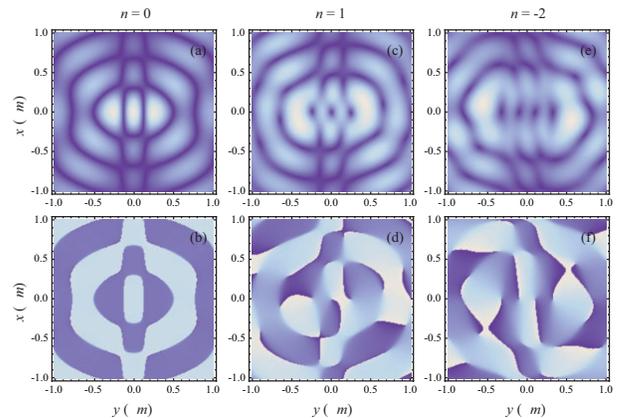}
\caption{Electric field $e_x (x,y)$ given in Eq.~(\ref{eq11}) for $L = 0.5\ {\mu m}$ and $\delta = 1$.
 For a given Bessel index $n$ the field amplitude is represented above and its complex argument below.} 
\label{fig06}
\end{figure}

As seen in the amplitude distribution of Figs.~\ref{fig04} and \ref{fig05}, the Bessel index $n$ induces a lateral shift of the twins in the presence of the thin film.
However, this is not a universal rule as shown in the following example.
The transverse profile of the electric field $e_x (x,y)$ for $L = 0.5\ {\mu m}$ and $\delta = 1$ is depicted in Fig.~\ref{fig06}.
In this case, the wave clearly penetrates into the core slab leading to a net transfer of energy from one side of the cladding to the other side.
Then the transmitted field is coupled with the counter-propagating incident wave in order to regenerate the Besselian pattern, which however shows a significant distortion and blurring.
Inside the core film, the transmitted fields are superposed with scattered waves originated from multiple reflections on the surfaces.
The multiple-beam interference leads to a standing wave in the $xy$ plane with resonant peaks and nodes (phase singularities) mainly along the $y$ axis.
For instance, three peaks form the shape of the diffraction-free beam if $n = 0$, as shown in Fig.~\ref{fig06}(a).
The profiles exhibit more complex features for higher $n$ orders.
In our numerical example, however, it gives rise to a pair of conspicuous spots of irregular contour if $n \neq 0$, seen in Figs.~\ref{fig06}(c) and \ref{fig06}(e).

\section{\label{sec04} Concluding remarks}

In this paper we have reported the formation of highly-localized nondiffracting beams in the presence of a core microfilm by means of a Besselian source.
Differential transverse patterns come out consisting mainly of a couple of non-centrosymmetric peaks enclosed by numerous phase singularities.
In general the index $n$ of the Bessel function fixes a shift, lying in a plane parallel to the interfaces, of the center of mass of each spot of light.
When the slab width is of the order of the wavelength, in principle, the neighboring twins may be indefinitely propagated brushing the plane surfaces.
Furthermore, a high concentration of electromagnetic energy density may be additionally found inside the monolayer if the relative increment of the dielectric constant is moderate or low.

Practical implementation of these ideas requires the same well-known procedures to produce a Bessel beam.
For instance, a field spectrum like that of Eq.~(\ref{eq07}) may be experimentally attained using an opaque screen, with a centered extremely-thin transparent annulus, placed at the front focal plane of a perfect lens \cite{Indebetouw89}. 
Highly-efficient approaches may be found using axicons \cite{Grunwald00,Zapata06e}, Fabry–-Perot interferometers \cite{Horvath97}, leaky screens \cite{Holm98,Reivelt02b,Reivelt02}, and diffractive optical elements \cite{Vasara89,Amako03,Li09}.
Using such devices as external sources in our system would excite the required diffraction-free wavefields in the layered medium.
Let us point out that the experimental conditions may impose dramatic limitations.
For instance, the generation of wave packets with mean propagation constant $\beta$ and on-axis spectral width $\Delta \beta$ leads to conservation of the beam intensity along a finite distance, $\Delta z \approx \lambda / \Delta \beta$ \cite{Indebetouw89}.

The theoretical analysis carried out in previous sections allows us to tackle more general problems in a plain way.
If the thin film is deposited on a substrate like glass, the ambient medium and the substrate constituting the cladding have different dielectric constants. 
The multilayered optical system is then described by incorporating two new wave fields $G_n^\pm$ for the substrate region, as those given in Eq.~(\ref{eq12}) for the core layer.
In this case we would use $k_{s,y} = \sqrt{k_s^2 - \beta^2 - k_x^2}$, where $k_s = \omega \sqrt{\epsilon_s} / c$ and $\epsilon_s$ is the dielectric constant of the substrate.
Finally the spectra $\tilde{F}_n^\pm$, $\tilde{G}_n^\pm$, and $\tilde{R}_n^\pm$ would be obtained from $\tilde{U}_n^\pm$ by applying appropriate boundary conditions.
Also, in practical implementations we might consider a $\pi$-illuminating configuration where counter-propagating fields do not impinge directly on the substrate.
Here we set $U_n^- = 0$ and wave localization is accomplished solely by the source field $U_n^+$ travelling in the ambient medium.
More sophisticated setups might be examined straightforwardly.

The properties and principles of Bessel-like beams in bulk media have offered multiple applications, including precision atomic micromanipulation \cite{Arlt00,Arlt00b,Grier03}, laser microfabrication of transparent materials \cite{Matsuoka06}, electron laser acceleration \cite{Hafizi97}, and highly-efficient generation of nonlinear effects \cite{Wulle93,Porras04}.
The formation of similar light structures near surfaces and interfaces leaves an open door to potential uses in non-destructive characterization of material thin films \cite{McGilp95}.  
Provided grazing incidence onto the core slab, such a procedure would increase wave penetration and enhance light-matter interaction.

\section*{Acknowledgments}

This research was funded by Ministerio de Ciencia e Innovaci\'on (MICIIN) under the project TEC2009-11635.

\bibliographystyle{osajnl}        % Include this if you use bibtex 

%%%%%%%%%%%%%%%%%%%%%%%%%%%%%%
%%Add figure captions%%%%%%%%%
%%%%%%%%%%%%%%%%%%%%%%%%%%%%%%
\newpage
\listoffigures

\end{document}